\documentclass[prl,aps,amssymb,twocolumn,superscriptaddress,notitlepage,longbibliography]{revtex4-2}
\usepackage[T1]{fontenc}
\usepackage{bbm}
\usepackage{amsmath}
\usepackage{amstext}
\usepackage{latexsym}
\usepackage{graphicx}
\usepackage{amsfonts}
\usepackage{epsfig}
\usepackage{dsfont}
\usepackage{mathrsfs}
\usepackage[usenames,dvipsnames]{xcolor}
\usepackage{hyperref}
\hypersetup{colorlinks=true, citecolor=MidnightBlue, urlcolor=MidnightBlue, linkcolor=MidnightBlue}
\usepackage{orcidlink}

\begin{document}

\title{Topological Defect Statistics in a Weakly First-Order Phase Transition}

\author{Antonin Rottman \orcidlink{0009-0005-1943-9317}}
\affiliation{Quantum Science and Engineering, Ecole Polytechnique F\'{e}d\'{e}rale de Lausanne (EPFL), Lausanne, Switzerland}

\author{Leonard Logari\'c \orcidlink{0000-0001-9486-2047}}
\affiliation{International Center for Elementary Particle Physics (ICEPP), University of Tokyo, 7-3-1 Hongo, Bunkyo-ku, Tokyo
113-0033, Japan}
\author{Fumika Suzuki \orcidlink{0000-0003-4982-5970}}\email{fsuzuki@icepp.s.u-tokyo.ac.jp}
\affiliation{International Center for Elementary Particle Physics (ICEPP), University of Tokyo, 7-3-1 Hongo, Bunkyo-ku, Tokyo
113-0033, Japan}

\author{Adolfo del Campo \orcidlink{0000-0003-2219-2851}}\email{adolfo.delcampo@uni.lu}
\affiliation{Department  of  Physics  and  Materials  Science,  University  of  Luxembourg,  L-1511  Luxembourg, Luxembourg}
\affiliation{Donostia International Physics Center,  E-20018 San Sebasti\'an, Spain}

\begin{abstract}
The celebrated Kibble-Zurek mechanism (KZM) characterizes the nonequilibrium dynamics of second-order phase transitions and predicts universal scaling of the average number of defects with the quench time. However, the KZM is not directly applicable to first-order phase transitions, where defect formation is instead governed by nucleation processes. Here, using a Ginzburg-Landau model, we investigate the defect-number statistics in a weakly first-order phase transition. We show that the resulting statistics contain signatures of both Kibble-Zurek dynamics and nucleation, revealing the interplay between these two defect-formation mechanisms. In particular, we find that the defect-number distribution evolves from approximately Gaussian to nearly Poissonian across the crossover from the KZM regime to nucleation-dominated defect formation. While the first two cumulants are well described by a simple additive estimate of the KZM and nucleation contributions, the third cumulant exhibits a plateau and an enhancement beyond this estimate, providing a sensitive probe of the interplay between the two defect-formation mechanisms.
\end{abstract}
\maketitle

\begin{figure*}
{%
\includegraphics[clip,width=1.9\columnwidth]{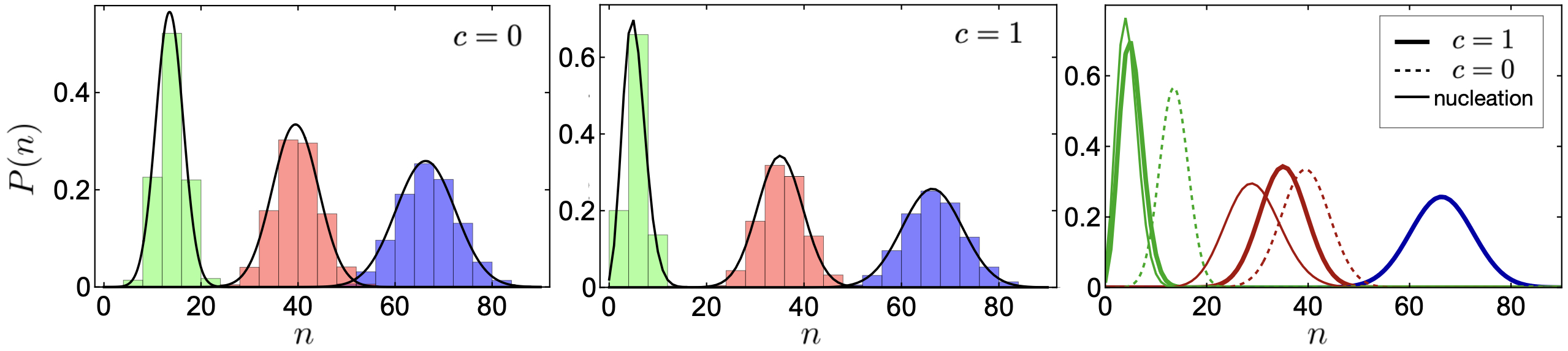} 
}
\caption{Dependence of the defect-number distribution $P(n)$ on the quench time $\tau_Q$ for $c=0$ (left) and $c=1$ (middle). The displayed distributions correspond to the quench timescales $\tau_Q=32$ (blue), $256$ (red), $8192$ (green). The right panel compares Gaussian fits for $c=1$ at $\tau_Q=32$ and $256$, a Poisson fit for $c=1$ at $\tau_Q=8192$ (thick line), a Gaussian fit for $c=0$ (dashed line), and a Poisson fit for nucleation (thin line). For $\tau_Q = 32$, the fit for $c = 1$ agrees with the fit for $c = 0$ (KZM). For $\tau_Q = 256$, the fit for $c=1$ shows contributions from both KZM and nucleation ($\epsilon^{*}=-0.31$). For $\tau_Q = 8192$, the fit for $c=1$ is in better agreement with the nucleation model ($\epsilon^{*}=-0.48$). For this histogram, we use $L=2048$ instead of $L=256$, as used in the other figures, to better resolve the shape of the defect-number distribution. } 
\label{fig1}
\end{figure*}

\begin{figure}
{%
\includegraphics[clip,width=1\columnwidth]{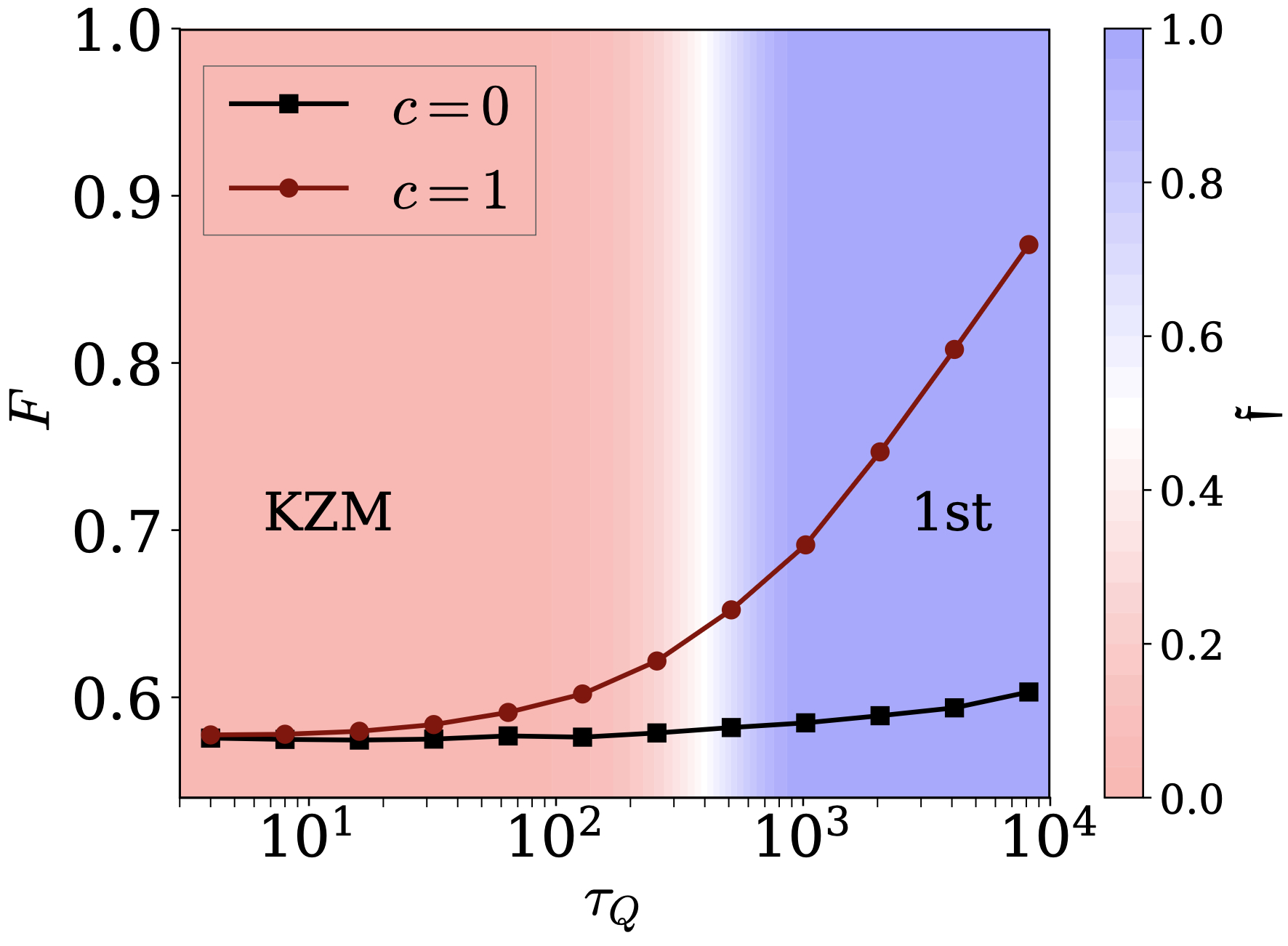} 
}
\caption{Dependence of the Fano factor $F=\kappa_2/\kappa_1$ on the quench time $\tau_Q$ for $c=0$ (black squares) and $c=1$ (red circles). The analytically estimated fraction of the system, $\mathfrak{f}$, occupied by the new phase due to nucleation events, as obtained from Eq.~(\ref{fraction}), is indicated by the background color scale.} 
\label{fig2}
\end{figure}

Phase transitions are ubiquitous in nature, governing phenomena ranging from structure formation in the early universe to condensed matter physics, with important applications in quantum technologies \cite{Dutta2015}. In particular,  second-order phase transitions are characterized by critical slowing down, i.e., the divergence of the relaxation time of the system in the neighborhood of the critical point. Consequently, crossing the critical point in a second-order phase transition is intrinsically nonadiabatic and results in the formation of topological defects. Near the adiabatic limit, under slow driving, the critical dynamics is described by the celebrated Kibble--Zurek mechanism (KZM) \cite{Kibble76a,Kibble76b,Zurek85,Zurek96c,DZ14}. The KZM predicts that the density of topological defects exhibits a universal power-law dependence as a function of the driving rate. The KZM has motivated a large body of theoretical work and has been experimentally verified in various physical systems, including trapped ions \cite{Ulm13,Pyka13}, colloidal monolayers \cite{Keim15}, ultracold gases \cite{Lamporesi2013,Ko2019,Goo21,Kim2023,clark2016universal}, multiferroics \cite{Lin14,Du2023}, quantum simulators \cite{Cui16,Keesling19}, and quantum annealing devices \cite{Weinberg20,Bando20,King22,King24}. 


The salient prediction of the KZM concerns the average density of defects. However, it has also been proposed that the entire defect-number distribution is universal \cite{delCampo2018Universal,GomezRuiz20,Balducci2023Large}. This implies that not only the average density, but also the fluctuations, are universal. Accordingly, all cumulants $\kappa_q$ ($q=1,2,\dots$) of the defect-number distribution exhibit a universal power-law scaling with the quench time. The universal power-law scaling of the variance and the third central moment of the defect-number distribution have been confirmed in numerical simulations of different models \cite{Mayo21,delCampo2021,Subires22,Thudiyangal24,Grabarits25,Xia2024lessons}. In addition, it has been experimentally verified using trapped-ion systems \cite{Cui20} and quantum simulators \cite{Bando20,King22,Kiss2025}. Deviations from this universality have been reported only in scenarios in which KZM breaks down. For example, the coupling to an environment generally induces heating and causes an increase in defect density with the quench time known as the anti-Kibble--Zurek mechanism \cite{Dutta16,Weinberg20,Miessen24}.  In such a scenario, the KZM breaks down and so does the universality of the defect-number statistics. Similarly, for systems with long-range interactions, the original formulation of the KZM is no longer applicable \cite{Caneva08,Jaschke17,Dutta17,Defenu18,Puebla20} and the defect-number statistics can deviate from the universal prediction \cite{Gherardini24,Grabarits25}. 

Despite these advances, progress in uncovering universal nonequilibrium dynamics beyond the KZM has largely been restricted to second-order phase transitions. Only recently have additional insights been obtained for weakly first-order transitions \cite{Suzuki24,Sadhasivam24, orona2026quantum, yang2026topological}. It is now understood that the average density of defects generated during a finite-time weakly first-order transition contains contributions associated with both KZM and nucleation. The statistical nature of defect-number fluctuations remains unexplored in weakly first-order transitions and is the focus of our work.\\

\textit{Model.---} To explore topological defect formation in a weakly first-order phase transition, we examine the numerical evolution of a one-dimensional system governed by the equation of motion for a real scalar field $\phi$. The equation is derived from the modified Ginzburg--Landau potential,
\begin{eqnarray}
V(\phi) = (\phi^4 - 2\epsilon \phi^2)/8 - c|\phi|^3/3,
\end{eqnarray}
where the first two terms account for the typical second-order phase transition behavior, whereas the third term introduces the first-order characteristics with $c>0$.

We assume that $\epsilon$ follows a linear quench, $\epsilon (t) =t/\tau_Q$ with $\tau_Q$ representing the quench timescale. 
The system is in contact with a thermal reservoir and
it obeys the Langevin equation,
\begin{eqnarray}\label{langevin}
\ddot{\phi}+\eta \dot{\phi}-\partial_{xx}\phi +\partial_{\phi} V(\phi) =\vartheta (x,t),
\end{eqnarray}
where  the noise term $\vartheta$ has correlations dictated by the fluctuation-dissipation theorem,
\begin{eqnarray}
\langle \vartheta (x,t) \vartheta  (x',t')\rangle = 2\eta \theta \delta (x'-x)\delta (t'-t).
\end{eqnarray}

The case $c=0$ and its extensions offer a paradigmatic testbed of KZM universality \cite{Laguna1997,Laguna1998,Dziarmaga1998,Dziarmaga1999,delcampo2010,Chiara2010,Power13,GomezRuiz20,Xia23prd,suzuki2025deconstructing,Suzuki2026machine}. Parity-symmetry breaking leads to the formation of $\mathbb{Z}_2$ kinks. The phase transition is characterized by mean-field critical exponents that take values $\nu=1/2$ and $z=2$ in the overdamped regime on which we focus. The KZM scaling law for the mean defect number $\kappa_1 {\propto}(\tau_0/\tau_Q)^{\frac{\nu}{1+z\nu}}$ then exhibits the power-law exponent $\nu/(1+z\nu)=1/4$. 

In what follows, we set $\eta=1$ and $\theta=0.01$. Unless stated otherwise, we use a system size of $L = 256$ with open boundary conditions and discretize it uniformly using $N = 512$ sites. We set the initial and final values of $\epsilon$ to $-2$ and $5$, respectively. The time evolution is implemented using the BAOAB integrator \cite{Leimkuhler2013}, with a time-step of $dt = 0.125$. The defects are counted as the number of sign changes in the field $\phi$.\\

\textit{Results.---} For finite $c$, the mean number of defects generated in a weakly first-order phase transition can be estimated as \cite{Suzuki24}
\begin{eqnarray}\label{kappa1}
\kappa_1= \mathfrak{f} \kappa_{1,\mathrm{nuc}} +(1-\mathfrak{f})  \kappa_{1,\mathrm{KZM}}
\end{eqnarray}
where $\kappa_{1,\mathrm{KZM}}$ is the first cumulant obeying the KZM in the second-order phase transition with $c=0$, while $\kappa_{1,\mathrm{nuc}}$ is the first cumulant arising from nucleation, which generally depends on the nucleation rate $\Gamma$ and the bubble-wall velocity $v$. The fraction of the system $\mathfrak{f}$ transformed into the new phase through a nucleation-and-growth process during the time interval $t\in [t_1,t_2]$ obeys the Avrami equation \cite{Avrami1939}:
\begin{eqnarray}\label{fraction}
\mathfrak{f}=1-\exp(-\Omega), \quad \Omega=\int_{t_1}^{t_2}\Gamma (\epsilon(t))\mathcal{V}(t,t_2)dt.
\end{eqnarray}
Here, $\mathcal{V}(t,t_2)=\int_t^{t_2} v(\epsilon(\tau))d\tau$ represents the volume of a nucleated bubble at time $t_2$, which was formed at time $t$. We use $t_1$, $t_2$, $v(\epsilon)=0.026\epsilon+0.016$ and $\Gamma(\epsilon(t))=0.4\exp[-B(\epsilon(t))/\theta]$, as obtained in Ref. \cite{Suzuki24} (see Appendix A). To estimate the nucleation contribution, we use the method proposed in Ref.~\cite{Suzuki24}. Since the nucleation barriers vary during the quench, we characterize the process by a representative value of the control parameter, $\epsilon^{*}=\epsilon(t^{*})$. The time $t^{*}$ is defined as the time at which the fraction of the system transformed by nucleation reaches half of its final value, $\mathfrak{f}^*=1-\exp\left(-\int_{t_1}^{t^*}\Gamma(\epsilon(t))\mathcal{V}(t,t^*)dt\right)=\mathfrak{f}/2$. We then approximate the first cumulant associated with nucleation by the value obtained for a time-independent system at $\epsilon^{*}$, i.e., $\kappa_{1,\mathrm{nuc}} =\kappa_{1,\mathrm{nuc}}(\epsilon^{*})$. This provides an effective estimate of the time-dependent nucleation contribution during the quench.

The important point of this estimate is that when the nucleation timescale is much longer than the quench timescale $\tau_Q$, there is insufficient time for nucleation to occur. Consequently, $\mathfrak{f}\rightarrow0$, and the KZM behavior is recovered, $\kappa_1 \simeq \kappa_{1,\mathrm{KZM}}$. Conversely, when the nucleation timescale becomes much shorter than the quench timescale, $\mathfrak{f}\rightarrow 1$, and the mean number of defects is governed by nucleation, $\kappa_1 \simeq \kappa_{1,\mathrm{nuc}}$.

Indeed, as shown in Fig.~\ref{fig1}, the defect-number distribution $P(n)$ for $c=1$ agrees with that of the KZM for $c=0$ when the quench timescale is short ($\tau_Q=32$). On the other hand, for long quench timescales ($\tau_Q=8192$), we find that the distribution for $c=1$ is in better agreement with the nucleation model ($\epsilon^*=-0.48$). For intermediate quench timescales ($\tau_Q=256$), the distribution for $c=1$ lies between the predictions of the KZM and nucleation models ($\epsilon^*=-0.31$). 

In general, when nucleation events \cite{turner1992bubble,ray2004measuring} are approximately independent, the number of nucleation-induced defects can be expected to closely follow a Poisson distribution , $P_\mathrm{Poisson}(n)=\frac{\lambda^n e^{-\lambda}}{n!},$ where $e$ is Euler's number and $\lambda$ is the mean number of  defects. In contrast, for a second-order phase transition governed by the KZM, the defect-number statistics can be described by a binomial or Poisson-binomial distribution, which approaches a Gaussian form in the large-system limit, $P_\mathrm{Gaussian}(n)\propto
\exp\left[-\frac{(n-\lambda)^2}{2\sigma^2}\right]$, where $\sigma^2$ is the variance. Although KZM defects can also approximately follow a Poisson distribution in the slow-quench regime, where only a few defects are produced, nucleation tends to yield a larger Fano factor than KZM defect formation for comparable mean defect numbers. 

\begin{figure}
{%
\includegraphics[clip,width=0.85\columnwidth]{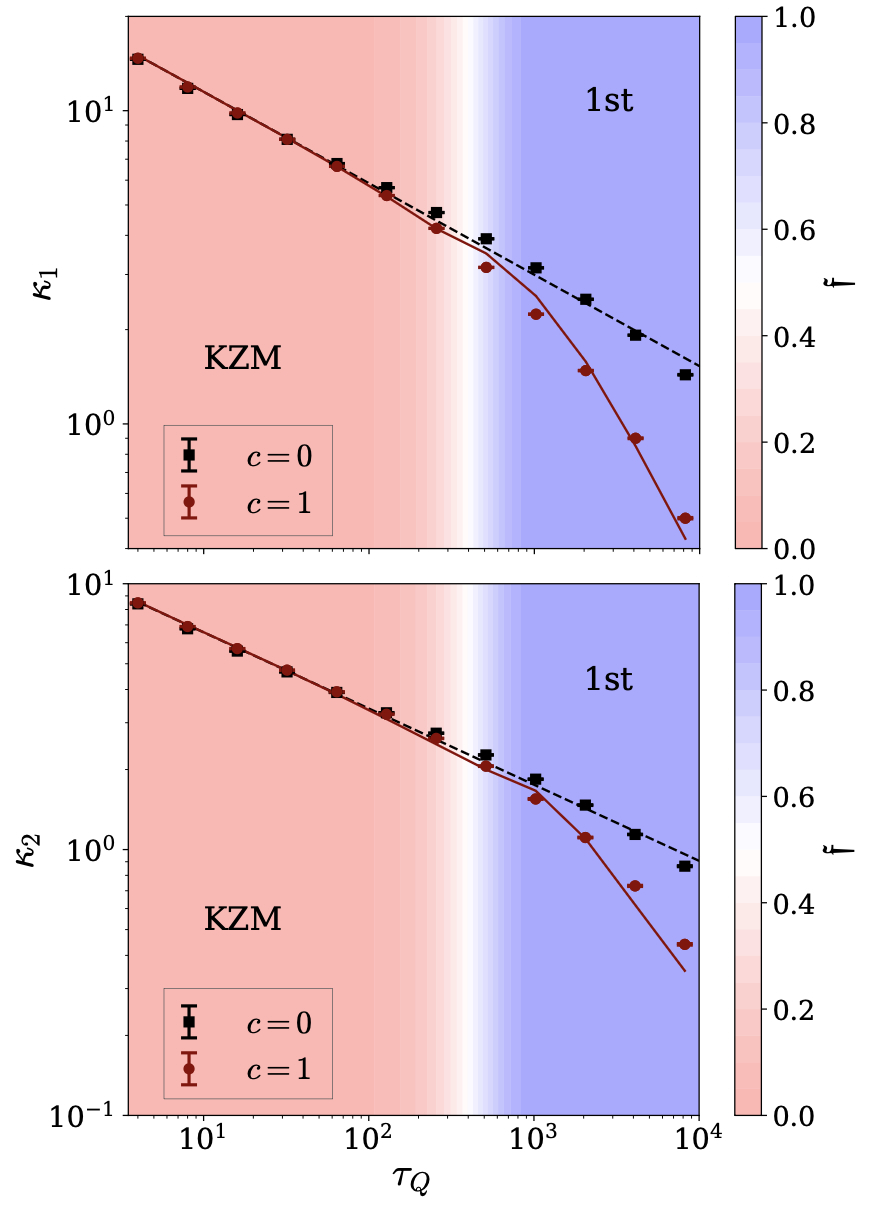} 
}
\caption{$\kappa_1$ and $\kappa_2$ as functions of the quench timescale $\tau_Q$.
Black squares represent the numerical results for $c=0$, where KZM power-law scaling holds, as shown by dashed black lines. The fits obey a power-law decay with an exponent of $-0.29$ for both $\kappa_1$ and $\kappa_2$, roughly consistent with the theoretical prediction $-\nu/(1+z\nu)=-1/4$. Red circles represent the numerical results for $c=1$. The solid  red lines show the analytical estimates obtained from Eqs.~(\ref{kappa1}) and (\ref{kappa2}), respectively. The analytically estimated fraction of the system, $\mathfrak{f}$, occupied by the new phase due to nucleation events, as obtained from Eq.~(\ref{fraction}), is indicated by the background color scale. Error bars are shown but are small.} 
\label{fig3}
\end{figure}

This interpretation is supported by Fig.~\ref{fig2}. The Fano factor $F=\kappa_2/\kappa_1$ for the second-order phase transition with $c=0$ remains approximately constant at $F\approx 0.58$ across all quench timescales, consistent with the KZM (a similar value was also reported in Ref. \cite{GomezRuiz20}). In contrast, for the weakly first-order phase transition with $c=1$, the Fano factor increases steadily from $F \approx 0.6$ to $F \approx 0.9$ as the quench timescale is varied from $\tau_Q \approx 10^2$ to $10^4$. These results for the defect statistics indicate that the nucleation is sub-Poissonian for the probed quench timescales and support the conjecture of Ref. \cite{Suzuki24} that, as the quench timescale increases, a system with weakly first-order characteristics crosses over from the KZM regime to a nucleation-dominated regime.

Although $\kappa_1$ is successfully predicted by Eq.~(\ref{kappa1}), an analytical prediction of higher-order cumulants can be  more challenging. The second cumulant, $\kappa_2=\langle (n-\langle n\rangle)^2\rangle$, measures the variance of the defect number and is therefore directly sensitive to fluctuations between different realizations, while the third cumulant, $\kappa_3=\langle (n-\langle n\rangle)^3\rangle$, additionally characterizes the asymmetry of these fluctuations and is particularly sensitive to the tails of the defect-number distribution. 

\begin{figure}
{%
\includegraphics[clip,width=0.9\columnwidth]{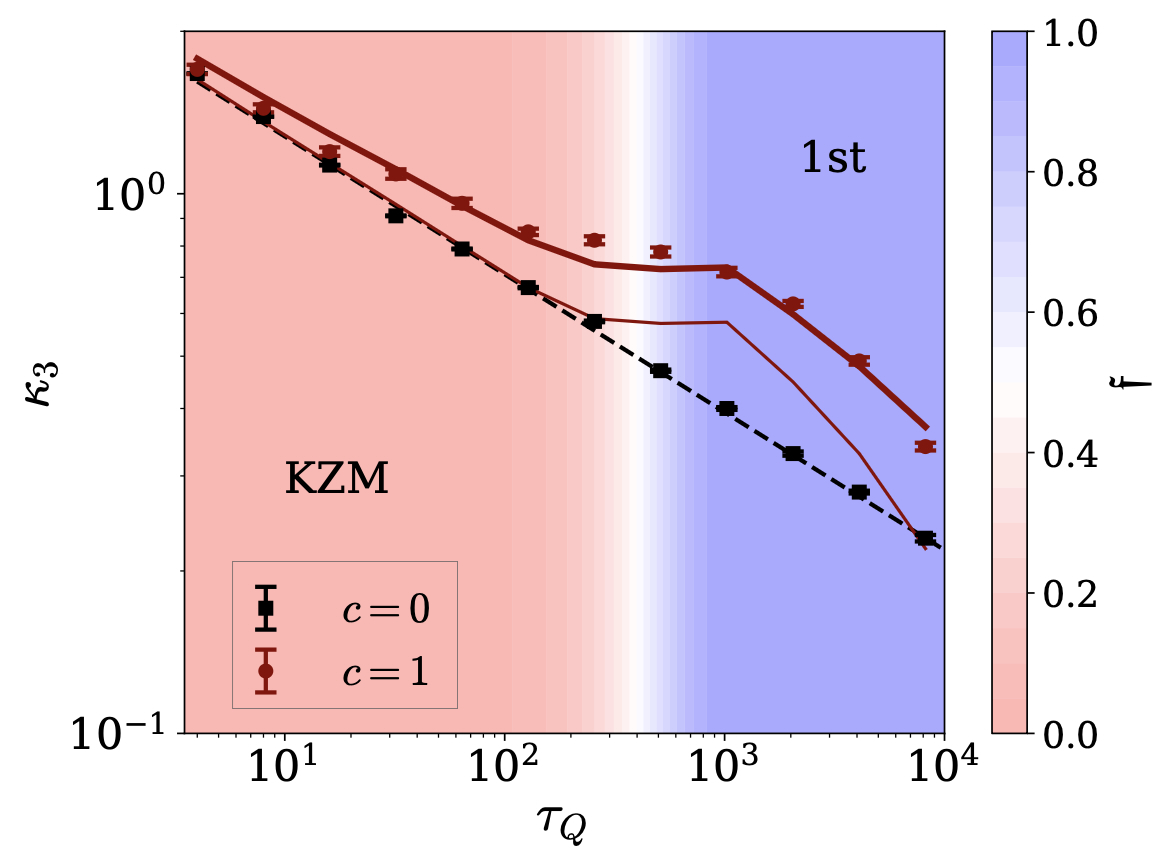} 
}
\caption{$\kappa_3$ as a function of the quench timescale $\tau_Q$. Black squares represent the numerical results for $c=0$, for which the KZM power-law scaling holds, as indicated by the dashed black line, with an exponent of $-0.26$. Red circles represent the numerical results for $c=1$. The solid red line shows the analytical estimate obtained from Eq.~(\ref{kappa3}), while the thick red line shows the corresponding estimate obtained by replacing $\kappa_{3,\mathrm{nuc}}$ and $\kappa_{3,\mathrm{KZM}}$ of Eq. (\ref{kappa3}) with the enhanced values $\tilde{\kappa}_{3,\mathrm{nuc}}$ and $\tilde{\kappa}_{3,\mathrm{KZM}}$, respectively, using $\Delta\kappa_{3,\mathrm{nuc}}=\Delta\kappa_{3,\mathrm{KZM}}=0.15$. The analytically estimated fraction of the system, $\mathfrak{f}$, occupied by the new phase due to nucleation events, as obtained from Eq.~(\ref{fraction}), is indicated by the background color scale. Error bars are shown but are small.} 
\label{fig4}
\end{figure}

Motivated by the picture of independent KZM and nucleation contributions, we assume that the cumulants are approximately additive over the fractions of the system governed by the two mechanisms (see Appendix B). This gives
\begin{eqnarray}\label{kappa2}
\kappa_2 \simeq
\mathfrak{f}\,\kappa_{2,\mathrm{nuc}}+(1-\mathfrak{f})\,\kappa_{2,\mathrm{KZM}}.
\end{eqnarray}

  Fig.~\ref{fig3} shows the numerical results for $\kappa_1$ and $\kappa_2$ as functions of the quench timescale $\tau_Q$. The black squares show the results for $c=0$, and the dashed black lines represent the corresponding fits. Both $\kappa_1$ and $\kappa_2$ obey the KZM power-law scaling. The red circles show the numerical results for the weakly first-order transition ($c=1$), while the solid red lines represent the analytical estimates given by Eqs.~(\ref{kappa1}) and (\ref{kappa2}) respectively, where $\kappa_{2,\mathrm{nuc}}=\kappa_{2,\mathrm{nuc}}(\epsilon^*)$. These estimates agree well with the numerical results, suggesting that the analytical formula provides an adequate description of the first two cumulants.

For the third cumulant, the corresponding relation can be written as
\begin{eqnarray}\label{kappa3}
\kappa_3 \simeq {}&\mathfrak{f}\,\kappa_{3,\mathrm{nuc}}+(1-\mathfrak{f})\,\kappa_{3,\mathrm{KZM}},
\end{eqnarray}
where $\kappa_{3,\mathrm{nuc}}=\kappa_{3,\mathrm{nuc}}(\epsilon^*)$.

Fig.~\ref{fig4} shows $\kappa_3$ as a function of the quench timescale $\tau_Q$. The red circles represent the numerical results for the weakly first-order phase transition ($c=1$), while the black squares for $c=0$ obey the KZM power-law scaling. For $c=1$, we observe a plateau in the crossover regime between KZM-dominated and nucleation-dominated defect formation. This behavior can be understood from the fact that the third cumulant of the  nucleation process, 
which is approximately Poissonian, is generally larger than that of the KZM process, whose defect-number distribution is close to Gaussian, even when their mean defect numbers, $\kappa_1$, are comparable. As the relative contribution of nucleation increases across the crossover, the larger third cumulant associated with nucleation compensates for the decrease with quench time, producing a plateau  in $\kappa_3$. Once the nucleation-dominated regime is reached, this compensation ceases and $\kappa_3$ decreases again. The solid red line represents the analytical estimate from Eq. (\ref{kappa3}). Although it exhibits behavior similar to the numerical result, including the plateau, a substantial deviation becomes apparent for the first time in the third cumulant. This indicates an additional contribution to the third-order fluctuations beyond the simple sum approximation of Eq. (\ref{kappa3}). The deviation in $\kappa_3$ suggests that the two defect-production mechanisms themselves may influence one another.

The observation suggests that $\kappa_3$ is larger than the KZM value even when $\mathfrak{f}$ is close to $0$, and larger than the estimate based on time-independent nucleation even when $\mathfrak{f}$ is close to $1$. In reality, nucleation is time dependent and can affect KZM domains, which may be key to understanding the enhancement of $\kappa_3$ in weakly first-order phase transitions. Even in the limit $\mathfrak{f}\rightarrow 0$, a small amount of nucleation can occur near the freeze-out time and locally disrupt domains formed through the KZM. If a nucleated region selects the same vacuum as the surrounding KZM domain, it has little effect. However, if it selects the opposite vacuum, the resulting interfaces may survive rather than annihilate, leading to an increased number of defects in rare realizations. Such rare events can generate a pronounced right tail in the defect-number distribution and therefore enhance $\kappa_3$. At the same time, because these events are rare, they can have only a small effect on $\kappa_1$ and $\kappa_2$ as see in Fig. \ref{fig3}. On the other hand, when $\mathfrak{f}$ is close to 1, it was sufficient to estimate $\kappa_{1,\mathrm{nuc}}$ and $\kappa_{2,\mathrm{nuc}}$ using the dominant nucleation contribution obtained from the time-independent-potential approximation. For $\kappa_{3,\mathrm{nuc}}$, however, the time dependence of the nucleation process appears to play an important role. As the potential evolves during the quench, the nucleation rate and number of defects generated by individual nucleation events vary over time. Rare nucleation events occurring at times when defect production is particularly efficient can therefore generate an enhanced high-defect-number tail. Such rare events may have only a small effect on the mean and variance, while contributing  to the third cumulant, thus producing a larger $\kappa_{3,\mathrm{nuc}}$ than predicted by the time-independent approximation.

To account for the possible enhancement of $\kappa_3$ through the process proposed above, we introduce $\tilde{\kappa}_{3,\mathrm{nuc}}=\kappa_{3,\mathrm{nuc}}+\Delta\kappa_{3,\mathrm{nuc}}$ and $\tilde{\kappa}_{3,\mathrm{KZM}}=\kappa_{3,\mathrm{KZM}}+\Delta\kappa_{3,\mathrm{KZM}}$, and estimate $\tilde{\kappa}_3 \simeq \mathfrak{f}\,\tilde{\kappa}_{3,\mathrm{nuc}}+(1-\mathfrak{f})\,\tilde{\kappa}_{3,\mathrm{KZM}}$, we obtain the thick red line in Fig. \ref{fig4} for $\Delta\kappa_{3,\mathrm{nuc}}=\Delta\kappa_{3,\mathrm{KZM}}=0.15$, which closely follows the numerical results.\\

\textit{Discussion and conclusion.---} In summary, we have characterized the full defect-number statistics across a weakly first-order phase transition driven in finite time. The distribution interpolates between the Gaussian statistics of the KZM at fast quenches and the nearly Poissonian statistics of independent nucleation events at slow quenches, with the Fano factor rising from $F \approx 0.58$ towards unity across the crossover. This change in statistics was also reflected in the third cumulant, $\kappa_3$, which develops a plateau in the intermediate regime. Such phenomenology is in stark contrast with continuous phase transitions, which exhibit uniform universal defect statistics at fast and slow quenches \cite{Xia2024lessons}.

We have seen that the conjecture proposed in~\cite{Suzuki24}, namely that the KZM and nucleation contributions can be combined according to the nucleated fraction $\mathfrak{f}$ determined from the Avrami equation, captures both the first and the second cumulants of the defect distribution. On the other hand, the third cumulant appears to receive additional contributions beyond the simple form of Eq.~(\ref{kappa3}). In particular, the modification of KZM-generated domains by nucleation events and the time dependence of the nucleation process may lead to an additional enhancement of $\kappa_3$ across the entire regime. The third cumulant thus provides a sensitive probe of the interplay between the two defect-formation mechanisms in a regime where the lower cumulants remain nearly indistinguishable from the independent-sum prediction given by Eqs. (\ref{kappa1}, \ref{kappa2}), without additional contributions arising from interactions or correlations between the two  mechanisms. Since defect-number distributions are routinely measured in trapped ions~\cite{Ulm13,Pyka13,Cui20}, ultracold gases~\cite{Lamporesi2013,Ko2019}, and quantum computing devices~\cite{Bando20,King22,Kiss2025,Visuri2026}, these signatures offer a concrete route to detect and quantify the interplay between Kibble-Zurek dynamics and nucleation in systems undergoing weakly first-order phase transitions.\\

{\it Acknowledgements.---} We thank Andr\'as Grabarits and Wojciech H. Zurek for helpful discussions. This work was supported in part by funding from Google.org. L.L. is supported by JST Grant Number JPMJPF2221 and IBM Quantum. AdC acknowledges financial support from the Luxembourg National Research Fund under Grant No. C22/MS/17132060/BeyondKZM.

\bibliography{references2.bib}

\appendix 
\section{Analytical Estimates of $t_1$, $t_2$, and  the Nucleation Rate}

For a weakly first-order transition, nucleation can be inferred to be significant over a finite time interval. We denote the beginning and end of this interval by $t_1$ and $t_2$, respectively. Here we derive the expressions for $t_1$ and $t_2$, as well as the nucleation rate $\Gamma$, following Ref.~\cite{Suzuki24}.

The potential is
\begin{equation}
V(\phi)=\frac{\phi^4-2\epsilon\phi^2}{8}-\frac{c}{3}|\phi|^3,
\end{equation}
with a linear quench $\epsilon(t)=\frac{t}{\tau_Q}$.

For $-c^2<\epsilon<0$, the metastable minimum at $\phi=0$ coexists
with the two broken-symmetry minima. The barriers separating these phases
are located at
\begin{equation}
\phi_{\rm barrier}=\pm (c-\sqrt{c^2+\epsilon}),
\end{equation}
with barrier height
\begin{equation}
h_{\rm barrier}=-\frac{1}{24}
\left(c-\sqrt{c^2+\epsilon}\right)^2\left[
3\epsilon+2c\left(c-\sqrt{c^2+\epsilon}\right)\right].
\end{equation}

Initially, the field fluctuates around the metastable state $\phi=0$.
Nucleation becomes possible once typical thermal fluctuations become large
enough to probe the barrier.

Near $\phi=0$, the potential is approximated by a harmonic potential,
\begin{equation}
V_{\rm har}(\phi)=\frac{1}{2}\omega^2\phi^2.
\end{equation}
The effective frequency $\omega$ is determined by requiring
\begin{equation}
V_{\rm har}(\phi_{\rm barrier})=h_{\rm barrier},
\end{equation}
so that
\begin{equation}
\omega^2=
\frac{2h_{\rm barrier}}{\phi_{\rm barrier}^2}.
\end{equation}

At temperature $\theta$, equipartition gives approximately
\begin{equation}
\langle \phi^2\rangle \simeq \frac{\theta}{\omega^2}.
\end{equation}

The time $t_1$ is then defined by the condition that the typical amplitude
of thermal fluctuations reaches the position of the nucleation barrier:
\begin{equation}
\sqrt{\langle \phi^2\rangle}=\phi_{\rm barrier}.
\end{equation}
Equivalently,
\begin{equation}
\sqrt{\frac{\theta}{\omega^2(\epsilon_1)}}=\phi_{\rm barrier}(\epsilon_1)
\end{equation}
where $\epsilon_1=\epsilon(t_1)$, and $t_1=\tau_Q\epsilon_1$.

Thus, $t_1$ marks the time at which fluctuations around the old phase
first become sufficiently large to interact appreciably with the
nucleation barrier.

The nucleation rate is written as
\begin{equation}
\Gamma(\epsilon(t))=A\exp\left[-\frac{B(\epsilon(t))}{\theta}\right],
\end{equation}
where $A=0.4$ for the parameter values used in this paper and the activation energy $B$ is
\begin{equation}
B(\epsilon)=2\int_0^{\phi_{\rm TP}}d\phi\,\sqrt{2V(\phi)}.
\end{equation}
Here, $\phi_{\rm TP}$ is the classical turning point that satisfies $V(\phi_{\rm TP})=V(0)=0$.

As $\epsilon\rightarrow0$, the barrier becomes progressively smaller, and
$B(\epsilon)\rightarrow0$. Eventually, the barrier energy becomes
comparable to the thermal kinetic energy of the field, characterized by
the temperature $\theta$.

The end of the nucleation-dominated interval is therefore defined by
\begin{equation}
B(\epsilon_2)=\theta,
\end{equation}
where $\epsilon_2=\epsilon(t_2)$ and $t_2=\tau_Q\epsilon_2$.

After $t_2$, the barrier is sufficiently small compared to the thermal
energy that the system no longer undergoes a conventional
barrier-crossing nucleation process. Instead, the field evolves toward
one of the broken-symmetry minima in a manner closer to the continuous
symmetry-breaking dynamics relevant to the Kibble-Zurek mechanism.

\section{Cumulants of the defect number}

We model the defect number as the sum of two statistically independent
contributions, $n = n_P + n_B$, produced by a Poisson mechanism acting on a
fraction $\mathfrak{f}$ of the system and by a mechanism of Bernoulli type acting
on the remaining fraction $1-\mathfrak{f}$. In the main text, $n_P=n_{\rm nuc}$ and
$n_B=n_{\rm KZM}$. In this idealized model, we take the nucleation contribution to
be Poissonian. We denote by $\kappa_{j,P}$ and $\kappa_{j,B}$ the cumulants of order
$j$ that each mechanism would produce if it acted on the entire system. The former
corresponds to a Poisson variable of mean $\lambda$, so that $\kappa_{j,P} = \lambda$
for all $j$, and the latter to a Poisson binomial variable $\sum_{i=1}^{M} B_i$ built
from $M$ independent Bernoulli variables with success probabilities $p_i$, which
becomes Gaussian only for large $M$.

Because the two contributions are independent, their cumulants are additive. Each
contribution scales with the fraction of the system on which it acts, provided
that the Poisson rate is uniform, that the $p_i$ are statistically equivalent
across the system, and that $\mathfrak{f}$ does not fluctuate between
realizations, as confirmed by our simulations. Under these conditions,
\begin{equation}
\kappa_j = \mathfrak{f}\,\kappa_{j,P} + (1-\mathfrak{f})\,\kappa_{j,B}
\label{eq:EM_additivity}
\end{equation}
holds for every order $j$. Note that Eq.~\eqref{eq:EM_additivity} describes a sum
of independent contributions, not a mixture of distributions (for the latter,
only the first cumulants  are additive).

Since the cumulants of a Bernoulli variable are polynomials in $p_i$, the power sums
$S_k = \sum_{i=1}^{M} p_i^k$ give
$\kappa_{1,B} = S_1$,
$\kappa_{2,B} = S_1 - S_2$, $\kappa_{3,B} = S_1 - 3S_2 + 2S_3$,
$\kappa_{4,B} = S_1 - 7S_2 + 12S_3 - 6S_4$, and so on; see,
e.g., \cite{delCampo2018Universal,Cui20}.

The Poisson contribution cancels in the differences of successive cumulants,
\begin{equation}
\begin{aligned}
\kappa_1 - \kappa_2 &= (1-\mathfrak{f})\textstyle\sum_i p_i^2, \\
\kappa_2 - \kappa_3 &= 2(1-\mathfrak{f})\textstyle\sum_i p_i^2 (1-p_i), \\
\kappa_3 - \kappa_4 &= 2(1-\mathfrak{f})\textstyle\sum_i p_i^2 (1-p_i)(2-3p_i),
\end{aligned}
\label{eq:EM_differences}
\end{equation}
so that cumulant differences probe the second mechanism alone.

The degree to which the two contributions can be separated depends on the knowledge of $p_i$, generally fixed by the geodesic rule \cite{GomezRuiz20}. If they are fixed by the underlying physics, the Poisson
contribution follows directly as $\mathfrak{f}\lambda = \kappa_1 - (1-\mathfrak{f})S_1$,
and the higher cumulants serve as consistency checks. For a homogeneous second
mechanism, $p_i = p$, the first three cumulants yield
\begin{align}
1 - p &= \frac{\kappa_2 - \kappa_3}{2(\kappa_1 - \kappa_2)}, \nonumber\\
(1-\mathfrak{f})M &= \frac{\kappa_1 - \kappa_2}{p^2}, \nonumber\\
\mathfrak{f}\lambda &= \kappa_1 - (1-\mathfrak{f})Mp ,
\label{eq:EM_binomial}
\end{align}
so that $\mathfrak{f}$ itself is determined once $\lambda$ or $M$ is known
independently. In the case of arbitrary $p_i$, the first three cumulants constrain
only $\kappa_1$, $(1-\mathfrak{f})S_2$, and $(1-\mathfrak{f})S_3$, and a parametric
description of the distribution of the $p_i$, or higher-order cumulants, is
required. In practice, cumulants are estimated using
$k$-statistics \cite{Fisher1930}, whose sampling variance grows rapidly with the
cumulant order, limiting the use of $\kappa_3$ and $\kappa_4$.

Approximating the Poisson binomial contribution by a Gaussian with matched mean and
variance sets $\kappa_{3,B}=0$, so that $\kappa_3 = \mathfrak{f}\lambda$, with an error
\begin{equation}
\delta\kappa_3 = (1-\mathfrak{f}) \sum_{i} p_i (1-p_i)(1-2p_i).
\label{eq:EM_error}
\end{equation}
This error is small when $p_i$ are close to $1/2$ or when nucleation dominates,
$|\delta\kappa_3| \ll \mathfrak{f}\lambda$. In the KZM regime, $p_i<1/2$, so the
Gaussian approximation overestimates $\mathfrak{f}\lambda$. Since $\delta\kappa_3$
grows with $M$, this bias persists even when the distribution itself looks Gaussian.
In the opposite limit $p_i \ll 1$, the Poisson binomial contribution becomes
Poissonian~\cite{LeCam1960,delCampo2018Universal,Cui20}, all cumulants approach
$\mathfrak{f}\lambda + (1-\mathfrak{f})S_1$, and the two mechanisms can no longer
be distinguished using cumulants.
\end{document}